\documentclass[letter,longauth]{aa} 

\usepackage{graphicx}
\usepackage{txfonts}
\usepackage[colorlinks=true, allcolors=blue]{hyperref}
\usepackage{orcidlink}
\usepackage{adjustbox}
\usepackage{multicol}
\usepackage{multirow}
\begin{document} 

   \title{Recovery of the X-ray polarisation of Swift~J1727.8$-$1613 after the soft-to-hard spectral transition}

   \titlerunning{Swift~J1727.8$-$1613 in the dim hard spectral state}

\author{J.~Podgorn{\'y} \inst{\ref{in:CAS-ASU}}\thanks{E-mail: jakub.podgorny@asu.cas.cz}\orcidlink{0000-0001-5418-291X}
\and J.~Svoboda \inst{\ref{in:CAS-ASU}}\orcidlink{0000-0003-2931-0742}
\and M.~Dov\v{c}iak\inst{\ref{in:CAS-ASU}}\orcidlink{0000-0003-0079-1239}    
\and A.~Veledina \inst{\ref{in:UTU},\ref{in:Nordita},\ref{in:IKI}}\orcidlink{0000-0002-5767-7253}
\and J.~Poutanen \inst{\ref{in:UTU},\ref{in:IKI}}\orcidlink{0000-0002-0983-0049}
\and P.~Kaaret \inst{\ref{in:MSFC}}\orcidlink{0000-0002-3638-0637}
\and S.~Bianchi \inst{\ref{in:UniRoma3}}\orcidlink{0000-0002-4622-4240}
\and A.~Ingram \inst{\ref{in:Newcastle}}\orcidlink{0000-0002-5311-9078}
\and F.~Capitanio \inst{\ref{in:INAF-IAPS}}\orcidlink{0000-0002-6384-3027}
\and S.~R.~Datta \inst{\ref{in:CAS-ASU}}\orcidlink{0000-0001-5975-1026}
\and E.~Egron \inst{\ref{in:Cagliari}}\orcidlink{0000-0002-1532-4142}
\and H.~Krawczynski \inst{\ref{in:Wash}}\orcidlink{0000-0002-1084-6507}
\and
G.~Matt \inst{\ref{in:UniRoma3}}\orcidlink{0000-0002-2152-0916}
\and F.~Muleri \inst{\ref{in:INAF-IAPS}}\orcidlink{0000-0003-3331-3794}
\and P.-O.~Petrucci \inst{\ref{in:Grenoble}}\orcidlink{0000-0001-6061-3480}
\and T.~D.~Russell \inst{\ref{in:Palermo} }\orcidlink{0000-0002-7930-2276}
\and J.~F.~Steiner \inst{\ref{in:CFA}}\orcidlink{0000-0002-5872-6061}
\and N.~Bollemeijer \inst{\ref{in:Amsterdam} }\orcidlink{0009-0005-6609-5852}
\and M.~Brigitte \inst{\ref{in:CAS-ASU},\ref{in:Charles}}\orcidlink{0009-0004-1197-5935}
\and N.~Castro Segura \inst{\ref{in:Warwick}}\orcidlink{0000-0002-5870-0443}
\and R.~Emami \inst{\ref{in:CFA}}\orcidlink{0000-0002-2791-5011}
\and J.~A.~Garc\'{i}a \inst{\ref{in:GSFC},\ref{in:Caltech}}\orcidlink{0000-0003-3828-2448}
\and K.~Hu \inst{\ref{in:Wash}}\orcidlink{0000-0002-9705-7948}
\and M.~N.~Iacolina\inst{\ref{in:ASI}}\orcidlink{0000-0003-4564-3416}
\and V.~Kravtsov \inst{\ref{in:UTU}}\orcidlink{0000-0002-7502-3173}
\and L.~Marra \inst{\ref{in:Padova},\ref{in:UniRoma3}}\orcidlink{0009-0001-4644-194X}
\and G.~Mastroserio \inst{\ref{in:Milano}}\orcidlink{0000-0003-4216-7936}
\and T.~Mu\~noz-Darias \inst{\ref{in:IAC},\ref{in:LaLaguna}}\orcidlink{0000-0002-3348-4035}
\and E.~Nathan \inst{\ref{in:Caltech}}\orcidlink{0000-0002-9633-9193}
\and M.~Negro \inst{\ref{in:Louisiana}}\orcidlink{0000-0002-6548-5622}
\and A.~Ratheesh \inst{\ref{in:INAF-IAPS}}\orcidlink{0000-0003-0411-4243}
\and N.~Rodriguez Cavero \inst{\ref{in:Wash}}\orcidlink{0000-0001-5256-0278}
\and R.~Taverna\inst{\ref{in:Padova}}\orcidlink{0000-0002-1768-618X}
\and F.~Tombesi \inst{\ref{in:UniRoma2},\ref{in:INFN-Roma2}}\orcidlink{0000-0002-6562-8654}
\and Y.~J.~Yang \inst{\ref{in:Taiwan},\ref{in:HongKong}}\orcidlink{0000-0001-9108-573X}
\and W.~Zhang \inst{\ref{in:NAO}}\orcidlink{0000-0003-1702-4917}
\and Y.~Zhang \inst{\ref{in:CFA},\ref{in:Groningen}}\orcidlink{0000-0002-2268-9318}
  }
  
  \authorrunning{J. Podgorn{\'y} et al.}
  
\institute{
Astronomical Institute of the Czech Academy of Sciences, Bo\v{c}n\'{i} II 1401/1, 14100 Praha 4, Czech Republic \label{in:CAS-ASU}
\and
Department of Physics and Astronomy, FI-20014 University of Turku, Finland \label{in:UTU}
\and
Nordita, KTH Royal Institute of Technology and Stockholm University, Hannes Alfv\'ens v\"ag 12, SE-10691 Stockholm, Sweden \label{in:Nordita}
\and
Space Research Institute, Russian Academy of Sciences, Profsoyuznaya 84/32, 117997 Moscow, Russia \label{in:IKI}
\and
NASA Marshall Space Flight Center, Huntsville, AL 35812, USA \label{in:MSFC} 
\and
Dipartimento di Matematica e Fisica, Universit\`{a} degli Studi Roma Tre, Via della Vasca Navale 84, 00146 Roma, Italy       \label{in:UniRoma3}     
\and 
School of Mathematics, Statistics, and Physics, Newcastle University, Newcastle upon Tyne NE1 7RU, UK   \label{in:Newcastle}          
\and
INAF Istituto di Astrofisica e Planetologia Spaziali, Via del Fosso del Cavaliere 100, 00133 Roma, Italy  \label{in:INAF-IAPS}          
\and
INAF Osservatorio Astronomico di Cagliari, Via della Scienza 5, 09047 Selargius (CA), Italy       \label{in:Cagliari} 
\and
Physics Department, McDonnell Center for the Space Sciences, and Center for Quantum Leaps, Washington University in St. Louis, St. Louis, MO 63130, USA \label{in:Wash} 
\and
Universit\'{e} Grenoble Alpes, CNRS, IPAG, 38000 Grenoble, France \label{in:Grenoble}
\and
INAF Istituto di Astrofisica Spaziale e Fisica Cosmica, Via U. La Malfa 153, I-90146 Palermo, Italy   \label{in:Palermo}          
\and
Center for Astrophysics $\vert$ Harvard \& Smithsonian, 60 Garden Street, Cambridge, MA 02138, USA  \label{in:CFA}       
\and
Anton Pannekoek Institute for Astronomy, Amsterdam, Science Park 904, NL-1098 NH, The Netherlands  \label{in:Amsterdam}          
\and
Astronomical Institute, Faculty of Mathematics and Physics, Charles University, V Holešovičkách 2, 18000 Prague 8, Czech Republic \label{in:Charles} 
\and
Department of Physics, University of Warwick, Gibbet Hill Road, Coventry CV4 7AL, UK
\label{in:Warwick}
\and
X-ray Astrophysics Laboratory, NASA Goddard Space Flight Center, Greenbelt, MD 20771, USA \label{in:GSFC} 
\and
California Institute of Technology, Pasadena, CA 91125, USA  \label{in:Caltech}     
\and
Agenzia Spaziale Italiana, Via della Scienza 5, 09047 Selargius (CA), Italy \label{in:ASI}
\and
Dipartimento di Fisica e Astronomia, Universit\`{a} degli Studi di Padova, Via Marzolo 8, 35131 Padova, Italy \label{in:Padova} 
\and
Dipartimento di Fisica, Universit\`{a} degli Studi di Milano, Via Celoria 16, I-20133 Milano, Italy \label{in:Milano} 
\and
Instituto de Astrof\'isica de Canarias, 38205 La Laguna, Tenerife, Spain \label{in:IAC}
\and
Departamento de Astrof\'\i{}sica, Universidad de La Laguna, E-38206 La Laguna, Tenerife, Spain   \label{in:LaLaguna}                
\and
Department of Physics and Astronomy, Louisiana State University, Baton Rouge, LA 70803, USA \label{in:Louisiana} 
\and
Physics Department, Tor Vergata University of Rome, Via della Ricerca Scientifica 1, 00133 Rome, Italy \label{in:UniRoma2}
\and
INFN - Rome Tor Vergata, Via della Ricerca Scientifica 1, 00133 Rome, Italy \label{in:INFN-Roma2}      
\and
Graduate Institute of Astronomy, National Central University, 300 Zhongda Road, Zhongli, Taoyuan 32001, Taiwan \label{in:Taiwan} 
\and
Laboratory for Space Research, The University of Hong Kong, Cyberport 4, Hong Kong            \label{in:HongKong} 
\and
National Astronomical Observatories, Chinese Academy of Sciences, 20A Datun Road, 100101 Beijing, China  \label{in:NAO}   
\and
Kapteyn Astronomical Institute, University of Groningen, P.O.\ BOX 800, 9700 AV Groningen, The Netherlands \label{in:Groningen} 
    }

   \date{Received ...; Accepted ...}

 
\abstract
{We report on the detection of X-ray polarisation in the black-hole X-ray binary Swift~J1727.8$-$1613 during its dim hard spectral state by the Imaging X-ray Polarimetry Explorer (IXPE). 
This is the first detection of X-ray polarisation at the transition from the soft to the hard state in an X-ray binary. 
We find an averaged 2--8 keV polarisation degree of $(3.3 \pm 0.4)\, \%$ and a corresponding polarisation angle of $3\degr\pm4\degr$, which matches the polarisation detected during the rising stage of the outburst, in September--October 2023, within $1\sigma$ uncertainty.
The observational campaign complements previous studies of this source and enables comparison of the X-ray polarisation properties of a single transient across the X-ray hardness-intensity diagram. The complete recovery of the X-ray polarisation properties, including the energy dependence, came after a dramatic drop in the X-ray polarisation during the soft state.
The new IXPE observations in the dim hard state at the reverse transition indicate that the accretion properties, including the geometry of the corona, appear to be strikingly similar to the bright hard state during the outburst rise despite the X-ray luminosities differing by two orders of magnitude. 
   }


   \maketitle
%

\section{Introduction}
\label{sec:intro}

Black-hole X-ray binary systems (BHXRBs) are unique cosmic laboratories for studying accretion physics in strong gravity. Particularly in the X-ray band, these systems are known to abruptly change their spectral and timing properties on timescales of weeks to months, implying changes in their innermost accretion configuration. In the X-ray hardness-intensity diagram, BHXRBs typically follow a hysteresis behaviour, resulting in a q-like curve that spans more than an order of magnitude in both directions with anti-clockwise evolution, often repeating multiple cycles \citep[e.g.][]{Remillard2006}.

In the `hard' state, the X-ray spectrum is dominated by a power law, originating from multiple Compton up-scatterings of low-energy disc or synchrotron seed photons in a hot optically thin corona or inner hot accretion flow \citep{Sunyaev1980, Zdziarski2004,Done2007,PoutanenVeledina2014}. 
Most of the known BHXRBs are transients, exhibiting a strong and sudden increase in X-ray flux in a hard spectral state, before transiting through `intermediate' states to the `soft' state \citep[e.g.][]{Belloni2010}. Such a transition occurs when there is a relative decrease in the hard power-law contribution with respect to the thermal emission from the accretion disc \citep{Shakura1973, Novikov1973}, the latter becoming dominant in the soft state. The hard state is usually associated with a significant accretion disc truncation up to a few tens of gravitational radii from the central stellar-mass black hole and with a presence of radio-emitting relativistic jets in polar directions \citep[e.g.][]{Fender2004,Done2007,Belloni2010}. Conversely, the soft state is characterised by a geometrically thin and optically thick accretion disc that extends to the innermost stable circular orbit and by quenched radio emission. After the X-ray flux decays, typically by about an order of magnitude from its peak outburst values, BHXRBs exhibit a reversed transition towards a power-law-dominated X-ray spectrum, and a re-launch of collimated bipolar radio ejecta is observed.
Finally, the `dim hard' state is reached, and sources decay towards quiescence.

\begin{figure*} 
\centering
  \begin{adjustbox}{valign=t}
\includegraphics[width=0.4\linewidth, trim={0.cm 0.2cm 0.cm 0.2cm}, clip]
{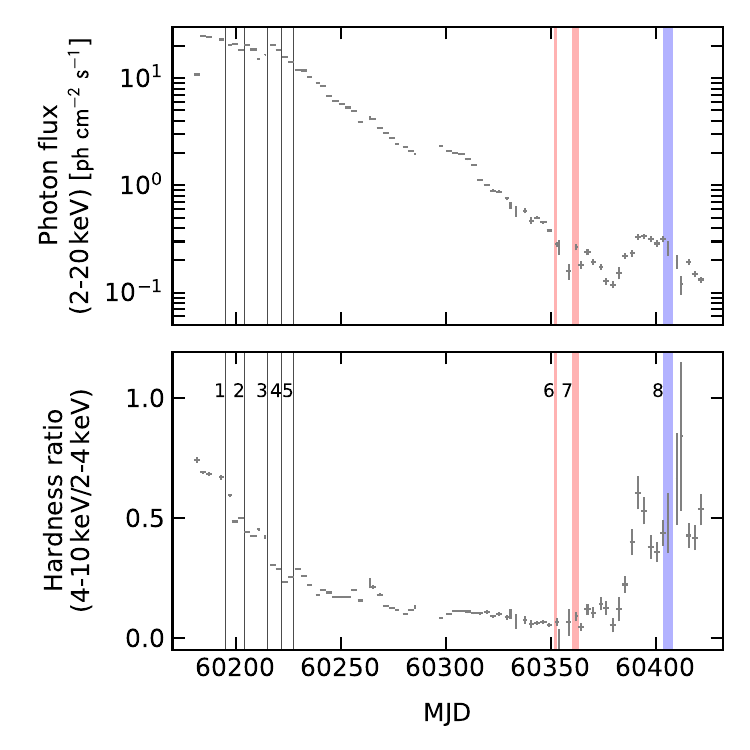}
\end{adjustbox}
  \begin{adjustbox}{valign=t}
\includegraphics[width=0.39\linewidth]{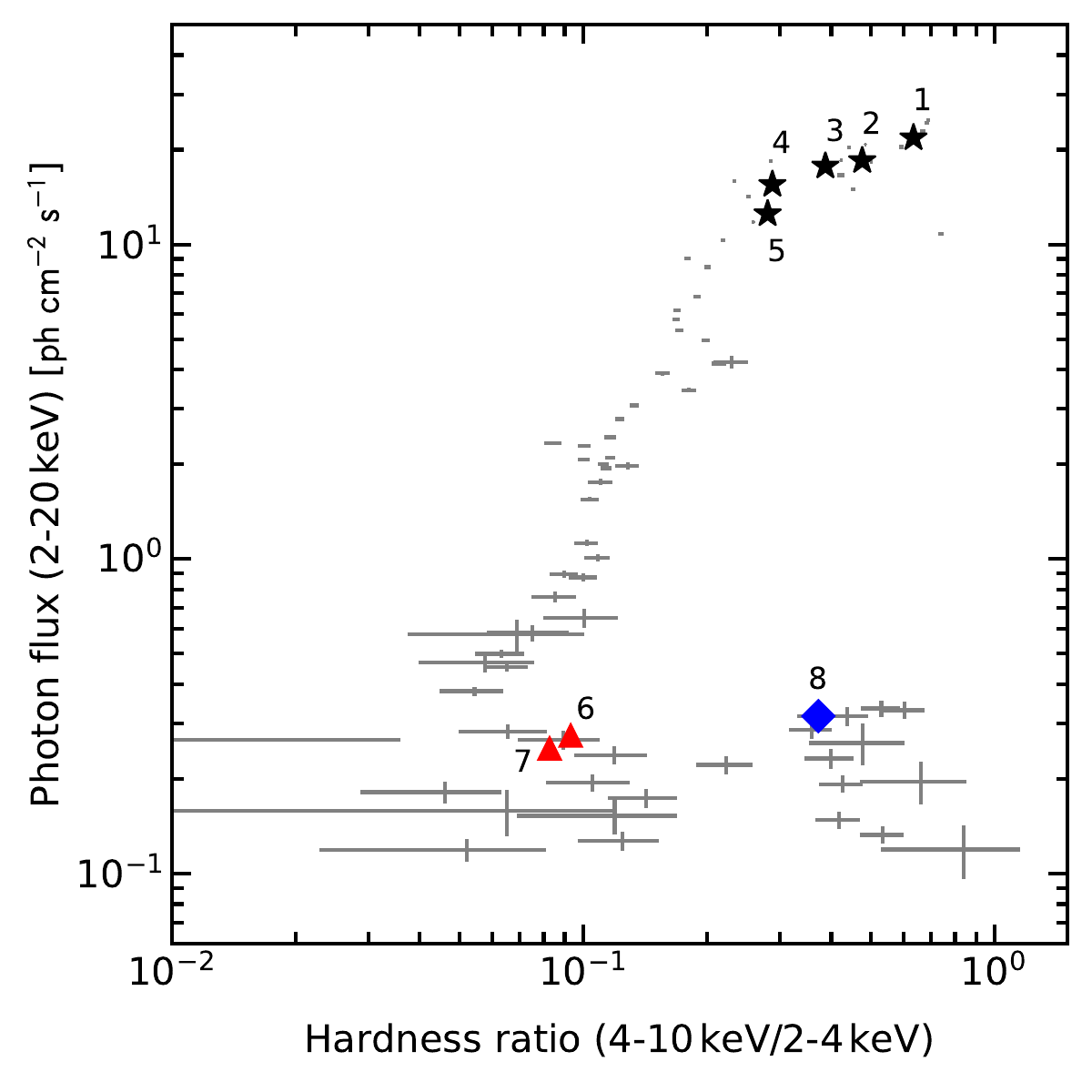}
\end{adjustbox}
\caption{Evolution of the flux and hardness ratio of Swift J1727.8$-$1613. 
The photon flux in the 2--20 keV range is shown in the upper-left panel. The hardness ratio, i.e. the ratio of the photon flux in the 4--10  and 2--4 keV bands, is shown in the lower-left panel. The fluxes are the 3-day averages from MAXI.  
The hardness-intensity diagram (right) goes from the beginning of the outburst until 2024 April 22. 
We indicate the initial five observations in the hard and intermediate states by IXPE \citep[described in detail in][]{Veledina2023, Ingram2024} as black stars, labelled 1--5. The two observations in the soft state \citep[described in detail in][]{Svoboda2024b} are marked by the red triangles, labelled 6 and 7. The new IXPE observation in the dim hard state, reported in this paper, is marked by the blue diamond symbol and labelled 8.} \label{fig:maxi}
\end{figure*}

In late August 2023, several teams reported a bright outburst ($\sim$7~Crab in the 2--20~keV energy range) of a new X-ray transient, Swift~J1727.8$-$1613  \citep{GCN.34540,GCN.34544}.
Subsequent X-ray spectroscopy \citep{ATel16210, ATel16217, Peng2024}, signatures of type-C quasi-periodic oscillations \citep[QPOs;][]{ATel16215, ATel16219, ATel16247, Zhao2024}, properties of the emitted radio jet \citep{ATel16211,ATel16228}, and optical spectroscopy \citep{Mata2024} all suggested a BHXRB classification. This was further confirmed by the anticipated transition through intermediate states towards the soft state over the next month \citep{Atel16289,Ingram2024}, which was accompanied by quenching of the QPOs \citep{ATel16273} and a subsequent decrease in the 2--20 keV flux by more than two orders of magnitude between October 2023 and February 2024 \citep{Svoboda2024b}. Monitor of All-sky X-ray Image \citep[MAXI;][]{Maxi_2009}  revealed that the source, while still displaying relatively high X-ray photon rates (0.2 ph\,s$^{-1}$\,cm$^{-2}$ in 2--20 keV),  began its transition towards the dim hard state around 2024 March 11 \citep{ATel16541}. The onset of the reverse transition was immediately confirmed by a corresponding radio detection of the compact jet \citep{ATel16552} with the Australia Telescope Compact Array.

 X-ray polarimetry is known for its potential to unravel the innermost accretion 
 geometry \citep[e.g.][]{Rees1975, Connors1977, SunyaevTitarchuk1985, Haardt1993, Matt1993,Poutanen1996,Dovciak2008}. The many theoretical studies can now be confronted with data from observations with the Imaging X-ray Polarimetry Explorer \citep[IXPE; launched in December 2021;][]{Weisskopf2022}, which enables 2--8 keV linear polarisation degree (PD) and polarisation angle (PA) measurements with remarkable precision. The BHXRB Cyg~X-1 was observed by IXPE in its hard state \citep[with a PD of $\sim$4\% in the 2--8 keV band;][]{Krawczynski2022} and its subsequent highly variable soft state \citep[PD of $\sim$2\% in 2--8 keV;][Steiner et al., in prep.]{ATel16084}. The PA in both periods was aligned with the observed projected radio jet directions within the uncertainties, indicating that a coronal plasma elongated in the accretion disc plane was the cause of the detected polarisation \citep[e.g.][]{Poutanen1996,Ursini2022,Krawczynski2022b}. Analogous results for X-ray polarisation from a Comptonised power law were also reported for the active galactic nuclei NGC~4151 \citep{Gianolli2023} and IC~4329A \citep{Ingram2023}.

Swift~J1727.8$-$1613 was first observed by IXPE in the hard to hard-intermediate state close to the outburst peak \citep{Veledina2023} and multiple times in the subsequent intermediate states \citep{Ingram2024}. The averaged 2--8 keV  PD decreased monotonically from $\sim$ 4\% to $\sim$ 3\%. 
The IXPE-measured PA was in all cases aligned with the radio, sub-millimetre, and optical PA \citep{Ingram2024,ATel16230,Atel16245}, indicating a similar alignment with the projected time-averaged jet orientation and similar coronal properties to the sources mentioned above. In all of the aforementioned IXPE polarisation observations, either the measured polarisation fraction was statistically favoured to increase with energy between 2--8 keV, with the associated PA consistent with being constant with energy, or the energy dependence of the polarisation was unconstrained. 

More recently, the source was also observed by IXPE during its soft state \citep{Svoboda2024b}. It was caught in an already decaying phase, with the X-ray flux about two orders of magnitude lower than during the peak of the outburst. Compared to the hard and intermediate states of Swift~J1727.8$-$1613, this observation revealed a substantial drop in the X-ray polarisation, showing only an averaged 2--8 keV  $1.2\%$ upper limit on the PD,
which indicates that the polarisation properties sensitively correspond to the changes in the spectral state.
The upper limit of the PD is consistent with the expectations of polarisation coming from emission from an optically thick accretion disc at low or intermediate inclination.

In this Letter we report on a new IXPE observation of Swift~J1727.8$-$1613, performed during its dim hard state just after the soft-to-hard transition. 
Comparison with the previous IXPE observations provides the until now missing X-ray polarimetric measurement at the end of the BHXRB's outburst, enabling us to compare X-ray polarimetric properties of a single source across the entire hardness-intensity diagram for the first time.

\section{Observations and data reduction}
\label{sec:data}

\subsection{MAXI}

The MAXI monitoring of Swift~J1727.8$-$1613  \citep{Maxi_2009} allowed us to characterise its broadband X-ray spectral properties on a daily basis. Using the MAXI `on-demand' service, in the same way as in \cite{Svoboda2024b}, we traced the evolution of the source on the hardness-intensity diagram. The results are shown in Fig.~\ref{fig:maxi} from the beginning of its outburst on 2023 August 24 until 2024 April 22. The position of the source during all previous IXPE observations,  1--7 \citep{Veledina2023, Ingram2024, Svoboda2024b}, and the new observation, 8, is indicated. The MAXI data confirm that IXPE newly observed Swift~J1727.8$-$1613 in the dim hard state, after which the source continued to decay towards quiescence. During the last IXPE exposure, the time-averaged X-ray luminosity was about two orders of magnitude lower than during its peak outburst, but the X-ray hardness was close to that in IXPE observation~3.

\subsection{IXPE}

The IXPE mission \citep{Weisskopf2022} operates with three detector units (DUs) on board. The reported new IXPE observation of Swift~J1727.8$-$1613 in the dim hard state  (ObsID: 03005801) was taken between UTC 15:46:55 on 2024 April 3 and UTC 02:24:41 on 2024 April 8, with a total live time of 202~ks. The Level-2 data were downloaded from the HEASARC archive. We used the {\tt xpselect} tool from the {\sc ixpeobssim} software \citep[version 31.0.1;][]{Baldini2022} to filter the source region, which was selected as a circle with a radius of 100\arcsec. We did not subtract the background, because the source was sufficiently bright \citep{diMarco2023}. The weighted Stokes $I$, $Q$, and $U$ spectra were obtained with the {\tt xselect} tool from the {\sc heasoft} package (version 6.33.1), using the command {\tt extract "SPECT" stokes=SIMPLE}. We used the {\sc heasoft} tool {\tt ixpecalcarf} on the event and attitude files of each DU ({\tt weight=2}, i.e. simple weighting) to work with the most up-to-date IXPE responses (v13). We did not re-bin the original IXPE data for the analysis below. We used the 2--8 keV energy range.

For the spectro-polarimetric analysis, we used {\sc xspec} \citep[version 12.14.0;][]{Arnaud1996}. All reported uncertainties are at the $68.3\%$ confidence level for one model parameter if not mentioned otherwise.

\section{Results}
\label{sec:results}

In this section we provide a time-integrated analysis of the IXPE data of Swift J1727.8--1613 across observation 8. The IXPE light curve is shown in Figure~\ref{fig:lightcurves}. The flux decays during the observation but no strong spectral variability is detected, and therefore, we proceeded with the analysis of the time-integrated spectra. 
We followed the steps for the spectro-polarimetric analysis 
that were used in \citet{Veledina2023}, \citet{Ingram2024}, and \citet{Svoboda2024b}. We performed a joint spectro-polarimetric fit of the Stokes $I$, $Q$, and $U$ spectra from all three DUs, tying the individual model parameters between the DUs except a cross-normalisation constant factor, {\tt const}, that accounts for the different absolute flux calibration of each DU. We used the {\tt tbabs} model \citep{Wilms2000} to account for line-of-sight absorption. We fixed the hydrogen column density, $N_\textrm{H} = 0.24 \times 10^{22} \, \mathrm{cm}^{-2}$, to the value obtained from the spectral fit of the soft-state X-ray spectrum \citep{Svoboda2024b}. Given the limited spectral capability of IXPE, we only assumed two major spectral components in the total 2--8 keV range. The thermal component is represented by the {\tt diskbb} model and contains a multi-temperature disc blackbody emission with peak temperature $kT_\mathrm{in}$ and a normalisation parameter. The Comptonised emission is represented by the {\tt powerlaw} model with a photon index $\Gamma$. The averaged 2--8 keV polarisation properties were obtained with a simple phenomenological model, {\tt polconst}, that assumes a constant PD and PA with energy in a given range. The full {\sc xspec} notation of the model is
\begin{equation}
    {\tt const \times tbabs  \times polconst  \times (diskbb + powerlaw)}\,.\\
    \label{model_polconst}
\end{equation}

\begin{figure} 
\centering
\includegraphics[width=0.8\linewidth]{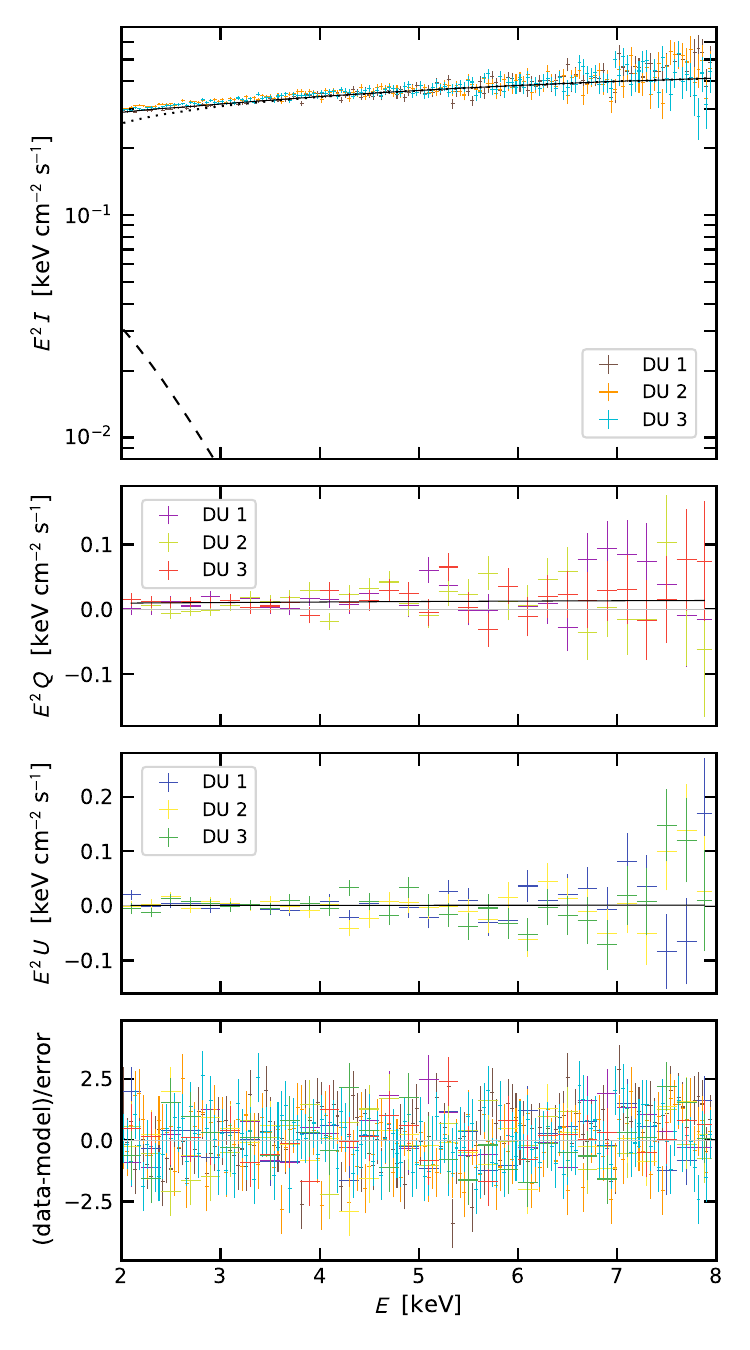}
\caption{X-ray time-averaged spectra of Swift J1727.8--1613 in the dim hard state (observation 8), as obtained from IXPE DU 1, DU 2, and DU 3. We show the unfolded Stokes parameters $I$, $Q$, and $U$, multiplied by $E^2$, alongside the best-fit Model (\ref{model_polconst}) from {\sc xspec}. The total model is shown with solid lines for all Stokes parameters. For Stokes $I$, we also display the {\tt diskbb} component (with dashed lines) and the {\tt powerlaw} component (with dotted lines) to illustrate that the contribution of the thermal component was very low in the 2--8 keV band, according to the IXPE data. The bottom panel shows the fit residuals. The $Q$ and $U$ data were re-binned by energy for plotting purposes only (using the {\sc xspec} command {\tt setplot rebin 10.0 5}).} \label{fig:spectra}
\end{figure}

\begin{figure*} 
\centering 
\includegraphics[width=0.29\linewidth, trim={0.5cm 0.cm 0.5cm 0.1cm}, clip]{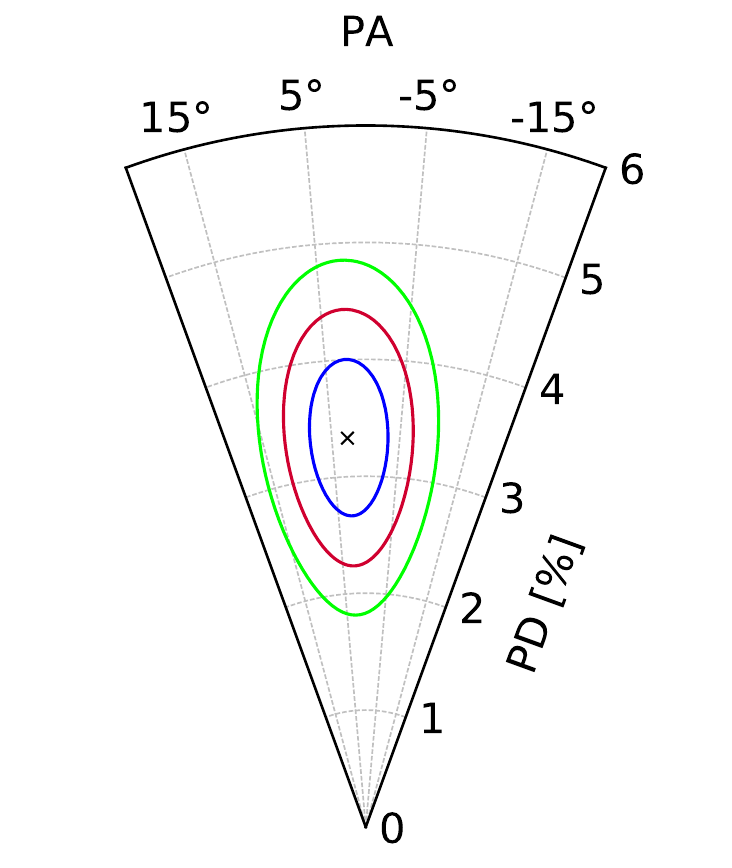}
\hspace{2cm}
\includegraphics[width=0.39\linewidth, trim={0cm 0.5cm 0cm 0.4cm}, clip]{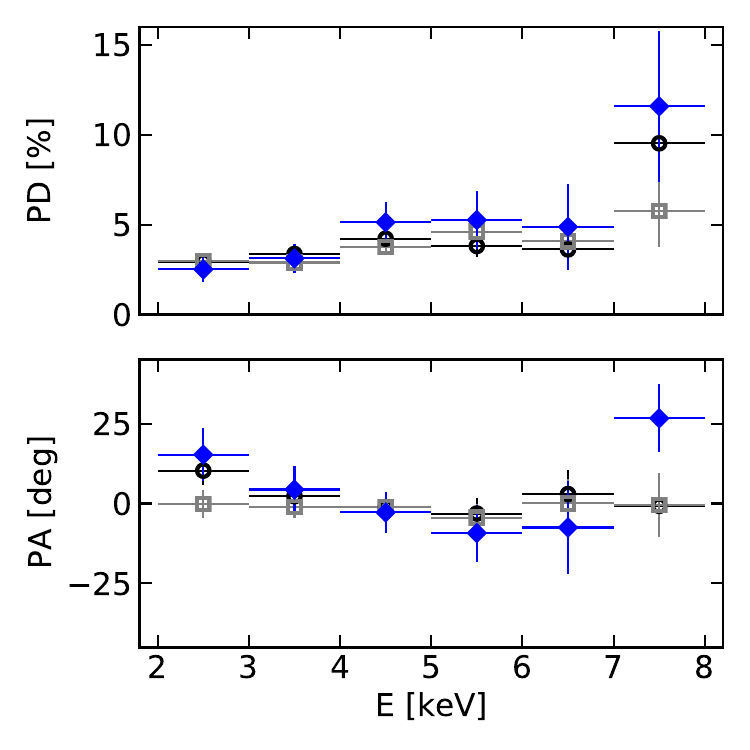}
\caption{Polarisation results from IXPE. Left: Averaged PD and PA in the 2--8 keV band, as obtained from {\sc xspec} with Model (\ref{model_polconst}) for observation~8, including the contours of the $68.3\%$ (blue), $95.5\%$ (red), and $99.7\%$ (green) confidence levels. The PA of $0\degr$ corresponds to the north direction, with positive values increasing eastwards. Right: PD and PA as a function of energy, as obtained from {\sc xspec} with Model (\ref{model_polconst}) for the same observation, 8 (blue diamonds). For comparison, we also show the energy dependence of polarisation for observations 3 (black circles) and 4 (grey squares) from \cite{Ingram2024}, obtained in a similar way. The uncertainties are given at the $68.3\%$ confidence level for one model parameter.} \label{fig:polconst}
\end{figure*}

Using the total 2--8 keV range, we obtained $\chi^2 / {\rm dof} = 1329/1333$. 
The best-fit spectral parameter values are listed in Table~\ref{t:spectralfitparams}. The unfolded Stokes $I$, $Q$, and $U$ spectra are shown in Fig.~\ref{fig:spectra}, along with the best-fit model and residuals. The spectrum is well described by a relatively hard power law with a photon index $\Gamma \approx 1.7$, with a very low contribution from the thermal component in the 2--8 keV range (<2\% in flux). 
For polarisation model {\tt polconst}, we get $\mathrm{PD} = (3.3 \pm 0.4)\, \%$ and $\mathrm{PA} = 3\degr \pm 4\degr$. Figure~\ref{fig:polconst} shows the obtained 2D contours from {\sc xspec} for different confidence levels between the 2--8 keV PD and PA in a polar plot.

\begin{table}
\begin{center}
\caption{Best-fit spectral parameter values of the joint IXPE spectro-polarimetric fit in {\sc xspec} with Model (\ref{model_polconst}). }
\resizebox{0.9\columnwidth}{!}{
\scriptsize
\begin{tabular}{p{1.5cm}p{2.3cm}p{1.9cm}}
\hline\hline
Component & Parameter [unit] & Value\\  
\hline 
{\tt tbabs} & $N_\textrm{H}$ [$10^{22}$\,cm$^{-2}$]  & $0.24$ (frozen) \\
\hline
{\tt diskbb} & $kT_\mathrm{in}$ [keV] &  $0.38\pm 0.04$ \\
    & norm [$10^2$] &  $6\substack{+7\\-3}$ \\
\hline
{\tt powerlaw} & $\Gamma$ & 
$1.73\pm 0.02$\\
& norm &  $0.239\substack{+0.007\\-0.008}$\\
\hline
{\tt const} & factor DU 1 & $1.0$ (frozen)\\
& factor DU 2 & $1.037\pm 0.003$\\
&  factor DU 3 & $1.022\pm 0.003$\\
\hline
\end{tabular}
}
\tablefoot{With Model (\ref{model_pollin}) we obtained the same best-fit values within the uncertainties.}
\label{t:spectralfitparams}
\end{center}
\end{table}

To examine the energy dependence of the detected X-ray polarisation, we fixed all parameter values from the best fit of Model (\ref{model_polconst}) at the 2--8 keV energy range except for the {\tt polconst} parameters, which were allowed to vary. With $\Delta E = 1 \, \mathrm{keV}$ energy binning, we fitted Model (\ref{model_polconst}) for each energy bin between 2 and 8 keV. The energy-dependent PD and PA are shown in Fig. \ref{fig:polconst} and compared to the previously reported energy-dependent results for observations 3 and 4 from \cite{Ingram2024}, obtained in a similar way.

To test whether the data from observation 8 favour a linear change in polarisation with energy over a constant-polarisation energy profile, we compared the best fit with Model (\ref{model_polconst}) at 2--8 keV with the same model, but using {\tt pollin} instead of {\tt polconst}:
\begin{equation}
    {\tt const  \times tbabs \times pollin  \times (diskbb + powerlaw)}\,.\\
    \label{model_pollin}
\end{equation}
The {\tt pollin} model has four parameters that set $\mathrm{PD}(E) = \mathrm{PD}_1+\mathrm{PD}_\mathrm{slope}(E-1\,\mathrm{keV})$ and $\mathrm{PA}(E) = \mathrm{PA}_1+\mathrm{PA}_\mathrm{slope}(E-1\,\mathrm{keV})$. We verified that letting $\mathrm{PA}_\mathrm{slope}$ vary does not significantly improve $\chi^2$ compared to freezing it to zero, and hence we froze $\mathrm{PA}_\mathrm{slope}$ to zero for the rest of the analysis. Following \cite{Ingram2024}, we defined a new parameter, $\mathrm{PD}_5 = \mathrm{PD}_1 + 4\mathrm{PD}_\mathrm{slope}$, so that the constant PD we fit is shifted to 5 keV (in the middle of the 2--8 keV IXPE bandpass) instead of the original 1 keV. We obtained $\chi^2 / {\rm dof} = 1323/1332$. The best-fit spectral parameter values did not vary from Model (\ref{model_polconst}) within the uncertainties. The best-fit polarisation values were $\mathrm{PD}_5 = (4.8 \pm 0.7)\, \%$, $\mathrm{PD}_\mathrm{slope} = (1.0 \pm 0.4)$\%~keV$^{-1}$, and $\mathrm{PA}_1 = 1\degr \pm 4\degr$. These results mean that the second fit with {\tt pollin} is only marginally preferred over the {\tt polconst} model. We performed an F test between the two best-fit solutions \citep[as in][]{Ingram2024} and conclude that the PD increases with energy at a $2.5\sigma$ confidence.

\section{Discussion}
\label{sec:discussion}

\begin{figure} 
\centering
\includegraphics[width=0.9\linewidth]{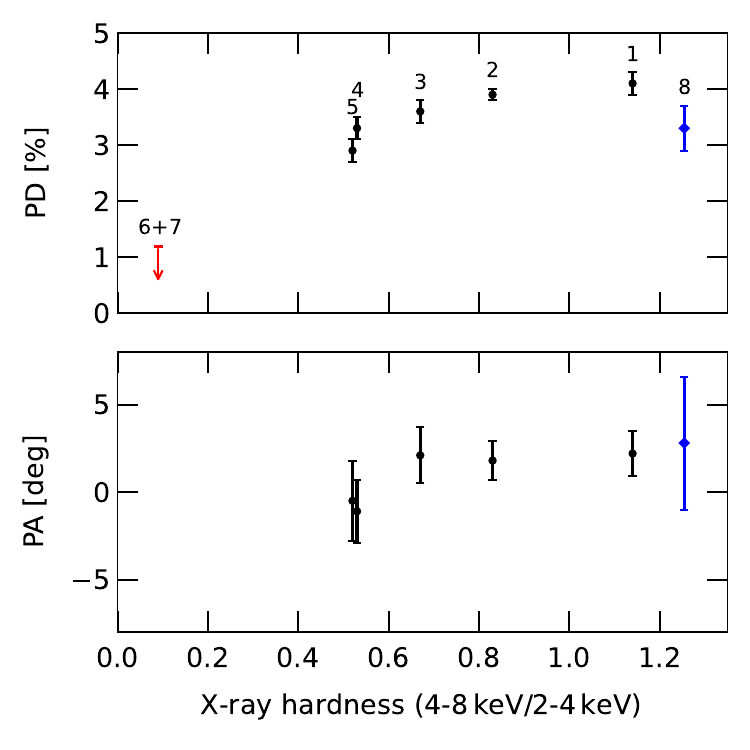}
\caption{PD and PA averaged in the 2--8 keV band as a function of X-ray hardness for all IXPE observations of Swift~J1727.8$-$1613, as measured by IXPE. This plot is an update of Fig.~5 from \cite{Svoboda2024b}. The red arrow signifies the 1.2\% upper limit on the PD (with the PA unconstrained) at the 99\% confidence level, as obtained in the soft state \citep[observations 6 and 7 combined;][]{Svoboda2024b}. Observation~8, of the dim hard state, is shown in blue. Observations 1--5, of the peak outburst \citep{Veledina2023, Ingram2024}, are shown in black.} \label{fig:pol_vs_hardness}
\end{figure}

Despite the source being two orders of magnitude less luminous during observation 8 than during the outburst peak, we observe the same X-ray polarisation properties in the hard (and hard-intermediate) state of Swift~J1727.8$-$1613 during the rising and decaying stages of the outburst, within 1$\sigma$ uncertainties \citep{Veledina2023, Ingram2024}. The energy-dependent values of PD and PA closely resemble the previous hard-state observations, showing a slight increase in PD with energy and a rather constant PA. Similar polarisation properties were also reported for the BHXRB Cyg~X-1 in the hard state \citep{Krawczynski2022} and for the nucleus of NGC 4151 \citep{Gianolli2023}, indicating a common accretion geometry. The new results suggest that the optically thin Comptonising medium in the decaying hard state of Swift~J1727.8$-$1613 is elongated in the accretion disc plane -- perpendicular to the presumed time-averaged radio jet direction in the rising hard state. Unfortunately, due to the relative radio faintness of the source during the reverse transition \citep{ATel16552}, radio polarisation measurements were not carried out \citep[the radio PD of a compact jet is expected to only be $<$1--3\%, i.e. below the sensitivity limits of the observations that were reported;][]{Fender2006}. Figure \ref{fig:pol_vs_hardness} shows the averaged 2--8 keV PD and PA of Swift~J1727.8$-$1613 versus X-ray hardness, as measured by IXPE from the beginning of its outburst, including the new observation. The source experienced a significant drop in X-ray polarisation fraction when reaching the soft state. This supports the basic deduction from X-ray spectroscopy that the physical properties of the accretion flow changed between the canonical hard and soft states \citep{Svoboda2024b}. 

BHXRB transients are known to follow the q-shape track in the hardness-intensity diagram \citep{Belloni2010}. An interesting feature is the presence of hysteresis in their path: the source attains the same hardness (corresponding to the same spectral slope) at different luminosities. 
For Swift J1727.8$-$1613, this difference reaches two orders of magnitude.
For the spectrum shaped by thermal Comptonisation, its spectral index is mostly determined by two parameters: the Thomson optical depth of the hot medium, $\tau_{\rm T}$, and the electron temperature, $kT_{\rm e}$ \citep[e.g.][]{Beloborodov99ASP}.  
Secondary effects are imposed by the characteristic energy of seed photons and the geometry. 
Here we considered the slab geometry, which gives the maximal PD for fixed other parameters.
For the observed two-orders-of-magnitude drop in luminosity in Swift J1727.7$-$1613, it is natural to expect a change in the parameters of the hot medium, while preserving the spectral slope.
Because the $\tau_{\rm T}$ of the hot medium is expected to drop linearly with the decrease in the mass accretion rate, $\dot M$ \citep[e.g. Eq.~(11) in][]{YuanNarayan2014}, the electron temperature has to increase to conserve the spectral slope; a lower scattering probability is compensated for by a higher energy shift in each scattering order. 
The decrease in $\tau_{\rm T}$ leads to an increase in PD for photons scattered many times in the medium \citep[e.g.][]{SunyaevTitarchuk1985}. 
On the other hand, an increase in $kT_{\rm e}$ induces stronger aberration effects, leading to a drop in the PD \citep{Poutanen1994}.  
This means that the spectral slope and the PD depend mostly on the same two parameters, and the mutual cancellation of their effect on both the spectral index and PD, if at all possible, may require some fine-tuning.

\begin{figure} 
\centering
\includegraphics[width=0.8\linewidth]{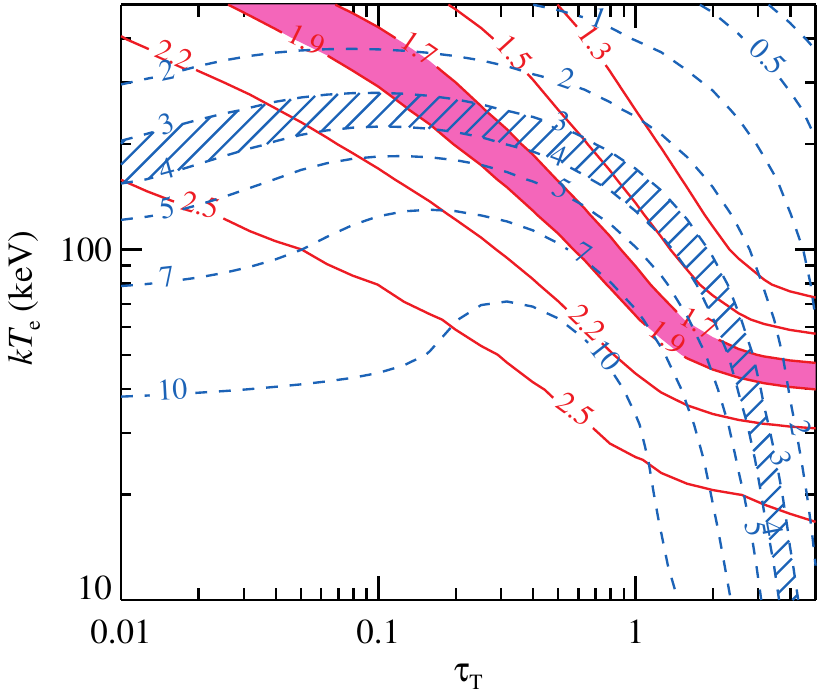}
\caption{Contour plot of constant photon index, $\Gamma$ (solid red), and the PD in~percent (dashed blue) on the plane electron temperature, $kT_{\rm e}$, versus Thomson optical depth, $\tau_{\rm T}$, computed for Comptonisation in a slab observed at inclination $i=45\degr$. 
The pink band represents the range $\Gamma=1.7$--1.9, which corresponds to the hard state spectra observed in Swift~J1727.8$-$1613. 
The blue hatched band corresponds to the PD of 3--4\% observed during observations 3, 4, and 8.
The two bands cross each other in two regions where $\tau_{\rm T}$ differs by an order of magnitude.} \label{fig:ssc_gamma_pd}
\end{figure}
 
To find the conditions whereby $\Gamma$ and the PD can be the same despite the orders-of-magnitude different luminosities, we computed a set of Comptonisation models in a slab geometry with the seed soft synchrotron photons peaking at 1~eV \citep{Veledina2013,PoutanenVeledina2014,YuanNarayan2014}. That could be relevant for 30 Schwarzschild radii, assuming a certain distribution of the magnetic field and optical depth in the hot accretion flow \citep[Eqs. 3 and 13 in][]{Veledina2013}, but the exact number has a minor effect on the resulting PD since in the IXPE range we see high scattering orders.
Thus, the only parameters of the problem are $kT_{\rm e}$  and $\tau_{\rm T}$.  
Simulations were done using the \texttt{compps} code \citep{Poutanen1996,veledina22,Poutanen2023}. 
The Comptonisation spectra observed at inclination $i=45\degr$ were fitted in the 2--8 keV energy band with a power law to determine $\Gamma$, and the PD was read at 4 keV.   
The contours of constant  $\Gamma$ and PD are shown in Fig.~\ref{fig:ssc_gamma_pd}. 
We see that there are indeed two solutions corresponding to the observed values of $\Gamma$ and PD; the $\tau_{\rm T}$ of the two solutions differ by a factor of 10. 
If the corresponding $\dot M$ also differs by the same factor, we have to conclude that the luminosity scales as ${\dot M}^2$, as is predicted for advection-dominated accretion flows \citep{NarayanYi95}. 
Our model also predicts that $kT_{\rm e}\sim250$~keV in the dim hard state, implying that no spectral cutoff should be observed up to $\sim$300~keV.  

We note that very similar polarisation signatures could be obtained for the case when seed photons are coming from the truncated disc, as long as the characteristic peak energy is considerably lower than the softest energy probed by IXPE \citep[see e.g. Fig. S9 in][]{Krawczynski2022}.
On the other hand, if the corona lies on top of the (non-truncated) disc, whose seed photon peak is close to the IXPE range \citep[e.g. as found in a fit to the bright hard state spectrum by][]{Peng2024}, the first Compton scattering order, which has polarisation perpendicular to the disc axis, would be clearly visible at $\sim$2--3~keV. 
Higher scattering orders are then polarised along the corona axis (in our case, the disc axis), and hence we expect to have a strong energy dependence of the PD in the IXPE range, potentially even a change in sign of the polarisation \citep[i.e. a rotation of the PA by 90\degr;][]{Poutanen1996,Krawczynski2022,Krawczynski2022b,Poutanen2023}, but neither of these signatures are observed.

We note that the simulations presented above account for neither general nor special relativistic effects related to the bulk motion of the gas in the close vicinity of a black hole, nor for the effects of non-local scattering.
Further observations, especially in the hard X-ray range, as well as more detailed modelling are needed to test our conclusions.

\begin{acknowledgements}
      We thank the IXPE operation team for their efforts in scheduling the observation. IXPE is a joint US and Italian mission. The US contribution is supported by the National Aeronautics and Space Administration (NASA) and led and managed by its Marshall Space Flight Center (MSFC), with industry partner Ball Aerospace (contract NNM15AA18C).  The Italian contribution is supported by the Italian Space Agency (Agenzia Spaziale Italiana, ASI) through contract ASI-OHBI-2022-13-I.0, agreements ASI-INAF-2022-19-HH.0 and ASI-INFN-2017.13-H0, and its Space Science Data Center (SSDC) with agreements ASI-INAF-2022-14-HH.0 and ASI-INFN 2021-43-HH.0, and by the Istituto Nazionale di Astrofisica (INAF) and the Istituto Nazionale di Fisica Nucleare (INFN) in Italy.  This research used data products provided by the IXPE Team (MSFC, SSDC, INAF, and INFN) and distributed with additional software tools by the High-Energy Astrophysics Science Archive Research Center (HEASARC), at NASA Goddard Space Flight Center (GSFC). This research has used the MAXI data provided by RIKEN, JAXA, and the MAXI team.
        
    J.Pod., J.S., M.D., and S.R.D. thank GACR project 21-06825X for the support and institutional support from RVO:67985815. 
    A.V. thanks the Academy of Finland grant 355672 and the Minobrnauki grant 23-075-67362-1-0409-000105 for support. Nordita is supported in part by NordForsk. S.B., F.C. and F.M. thank the INAF grant 1.05.23.05.06: “Spin and Geometry in accreting X-ray binaries: The first multi frequency spectro-polarimetric campaign”. A.I. acknowledges support from the Royal Society. The work of G.Matt, F.M., L.M., and R.T. is partially supported by the PRIN grant 2022LWPEXW of the Italian Ministry of University and Research (MUR). P.-O.P. thanks the French Space Agency (CNES). 
    M.B. acknowledges the support from GAUK project No. 102323. V.K. acknowledges support from the Finnish Cultural Foundation. G.Mas. acknowledges financial support from the European Union's Horizon Europe research and innovation programme under the Marie Sk\l{}odowska-Curie grant agreement No. 101107057. T.M-D. acknowledges support from the Spanish \textit{Agencia estatal de investigaci\'on} via PID2021-124879NB-I00. Y.Z. acknowledges support from the Dutch Research Council (NWO) Rubicon Fellowship, file No. 019.231EN.021.

\end{acknowledgements}

\bibliography{refs}

\begin{thebibliography}{}
\providecommand\natexlab[1]{#1}
\providecommand\JournalTitle[1]{#1}

\bibitem[{{Arnaud}(1996)}]{Arnaud1996}
{Arnaud}, K.~A. 1996, in ASP Conf. Ser., Vol. 101, Astronomical Data Analysis Software and Systems V, ed. G.~H. {Jacoby} \& J.~{Barnes} (San Francisco: ASP), 17

\bibitem[{{Baldini} {et~al.}(2022){Baldini}, {Bucciantini}, {Lalla}, {Ehlert}, {Manfreda}, {Negro}, {Omodei}, {Pesce-Rollins}, {Sgr{\`o}}, \& {Silvestri}}]{Baldini2022}
{Baldini}, L., {Bucciantini}, N., {Lalla}, N.~D., {et~al.} 2022, \href{http://dx.doi.org/10.1016/j.softx.2022.101194}{\JournalTitle{SoftwareX}, 19, 101194}

\bibitem[{{Belloni}(2010)}]{Belloni2010}
{Belloni}, T.~M. 2010, \href{http://dx.doi.org/10.1007/978-3-540-76937-8_3}{in Lecture Notes in Physics, Vol. 794, The Jet Paradigm, ed. T.~{Belloni}} (Berlin: Springer Verlag), 53

\bibitem[{{Beloborodov}(1999)}]{Beloborodov99ASP}
{Beloborodov}, A.~M. 1999, \href{http://dx.doi.org/10.48550/arXiv.astro-ph/9901108}{in ASP Conf. Ser., Vol. 161, High Energy Processes in Accreting Black Holes, ed. J.~{Poutanen} \& R.~{Svensson}} (San Francisco: ASP), 295

\bibitem[{{Bollemeijer} {et~al.}(2023{\natexlab{a}}){Bollemeijer}, {Uttley}, {Buisson}, {Homan}, {Altamirano}, {Gendreau}, {Arzoumanian}, {Strohmayer}, \& {Sanna}}]{ATel16247}
{Bollemeijer}, N., {Uttley}, P., {Buisson}, D., {et~al.} 2023{\natexlab{a}}, \JournalTitle{The Astronomer's Telegram}, 16247, 1

\bibitem[{{Bollemeijer} {et~al.}(2023{\natexlab{b}}){Bollemeijer}, {Uttley}, {Buisson}, {Homan}, {Altamirano}, {Gendreau}, {Arzoumanian}, {Strohmayer}, \& {Sanna}}]{ATel16273}
---. 2023{\natexlab{b}}, \JournalTitle{The Astronomer's Telegram}, 16273, 1

\bibitem[{{Bright} {et~al.}(2023){Bright}, {Farah}, {Fender}, {Siemion}, {Pollak}, \& {DeBoer}}]{ATel16228}
{Bright}, J., {Farah}, W., {Fender}, R., {et~al.} 2023, \JournalTitle{The Astronomer's Telegram}, 16228, 1

\bibitem[{Connors \& Stark(1977)}]{Connors1977}
Connors, P.~A., \& Stark, R.~F. 1977, \href{http://dx.doi.org/10.1038/269128a0}{\JournalTitle{Nature}, 269, 128}

\bibitem[{{Di Marco} {et~al.}(2023){Di Marco}, {Soffitta}, {Costa}, {Ferrazzoli}, {La Monaca}, {Rankin}, {Ratheesh}, {Xie}, {Baldini}, {Del Monte}, {Ehlert}, {Fabiani}, {Kim}, {Muleri}, {O'Dell}, {Ramsey}, {Rubini}, {Sgr{\`o}}, {Silvestri}, {Tennant}, \& {Weisskopf}}]{diMarco2023}
{Di Marco}, A., {Soffitta}, P., {Costa}, E., {et~al.} 2023, \href{http://dx.doi.org/10.3847/1538-3881/acba0f}{\JournalTitle{\aj}, 165, 143}

\bibitem[{{Done} {et~al.}(2007){Done}, {Gierli{\'n}ski}, \& {Kubota}}]{Done2007}
{Done}, C., {Gierli{\'n}ski}, M., \& {Kubota}, A. 2007, \href{http://dx.doi.org/10.1007/s00159-007-0006-1}{\JournalTitle{\aapr}, 15, 1}

\bibitem[{{Dov{\v{c}}iak} {et~al.}(2008){Dov{\v{c}}iak}, {Muleri}, {Goosmann}, {Karas}, \& {Matt}}]{Dovciak2008}
{Dov{\v{c}}iak}, M., {Muleri}, F., {Goosmann}, R.~W., {Karas}, V., \& {Matt}, G. 2008, \href{http://dx.doi.org/10.1111/j.1365-2966.2008.13872.x}{\JournalTitle{\mnras}, 391, 32}

\bibitem[{{Dovčiak} {et~al.}(2023){Dovčiak}, {Steiner}, {Krawczynski}, \& {Svoboda}}]{ATel16084}
{Dovčiak}, M., {Steiner}, J.~F., {Krawczynski}, H., \& {Svoboda}, J. 2023, \JournalTitle{The Astronomer's Telegram}, 16084, 1

\bibitem[{{Draghis} {et~al.}(2023){Draghis}, {Miller}, {Homan}, {Uttley}, {Bollemeijer}, {Steiner}, {Hare}, {Tombesi}, {Gendreau}, {Arzoumanian}, {Strohmayer}, {Sanna}, {Altamirano}, {Buisson}, \& {Fabian}}]{ATel16219}
{Draghis}, P.~A., {Miller}, J.~M., {Homan}, J., {et~al.} 2023, \JournalTitle{The Astronomer's Telegram}, 16219, 1

\bibitem[{{Fender}(2006)}]{Fender2006}
{Fender}, R. 2006, in Cambridge Astrophysics Series, Vol.~39, Compact stellar X-ray sources, ed. W.~{Lewin} \& M.~{van der Klis} (Cambridge: Cambridge University Press), 381

\bibitem[{{Fender} {et~al.}(2004){Fender}, {Belloni}, \& {Gallo}}]{Fender2004}
{Fender}, R.~P., {Belloni}, T.~M., \& {Gallo}, E. 2004, \href{http://dx.doi.org/10.1111/j.1365-2966.2004.08384.x}{\JournalTitle{\mnras}, 355, 1105}

\bibitem[{{Gianolli} {et~al.}(2023){Gianolli}, {Kim}, {Bianchi}, {Ag{\'\i}s-Gonz{\'a}lez}, {Madejski}, {Marin}, {Marinucci}, {Matt}, {Middei}, {Petrucci}, {Soffitta}, {Tagliacozzo}, {Tombesi}, {Ursini}, {Barnouin}, {De Rosa}, {Di Gesu}, {Ingram}, {Loktev}, {Panagiotou}, {Podgorny}, {Poutanen}, {Puccetti}, {Ratheesh}, {Veledina}, {Zhang}, {Agudo}, {Antonelli}, {Bachetti}, {Baldini}, {Baumgartner}, {Bellazzini}, {Bongiorno}, {Bonino}, {Brez}, {Bucciantini}, {Capitanio}, {Castellano}, {Cavazzuti}, {Chen}, {Ciprini}, {Costa}, {Del Monte}, {Di Lalla}, {Di Marco}, {Donnarumma}, {Doroshenko}, {Dov{\v{c}}iak}, {Ehlert}, {Enoto}, {Evangelista}, {Fabiani}, {Ferrazzoli}, {Garc{\'\i}a}, {Gunji}, {Heyl}, {Iwakiri}, {Jorstad}, {Kaaret}, {Karas}, {Kislat}, {Kitaguchi}, {Kolodziejczak}, {Krawczynski}, {La Monaca}, {Latronico}, {Liodakis}, {Maldera}, {Manfreda}, {Marscher}, {Marshall}, {Massaro}, {Mitsuishi}, {Mizuno}, {Muleri}, {Negro}, {Ng}, {O'Dell}, {Omodei}, {Oppedisano}, {Papitto}, {Pavlov}, {Peirson}, {Perri},
  {Pesce-Rollins}, {Pilia}, {Possenti}, {Ramsey}, {Rankin}, {Roberts}, {Romani}, {Sgr{\`o}}, {Slane}, {Spandre}, {Swartz}, {Tamagawa}, {Tavecchio}, {Taverna}, {Tawara}, {Tennant}, {Thomas}, {Trois}, {Tsygankov}, {Turolla}, {Vink}, {Weisskopf}, {Wu}, {Xie}, \& {Zane}}]{Gianolli2023}
{Gianolli}, V.~E., {Kim}, D.~E., {Bianchi}, S., {et~al.} 2023, \href{http://dx.doi.org/10.1093/mnras/stad1697}{\JournalTitle{\mnras}, 523, 4468}

\bibitem[{{Haardt} \& {Matt}(1993)}]{Haardt1993}
{Haardt}, F., \& {Matt}, G. 1993, \href{http://dx.doi.org/10.1093/mnras/261.2.346}{\JournalTitle{\mnras}, 261, 346}

\bibitem[{{Ingram} {et~al.}(2023){Ingram}, {Ewing}, {Marinucci}, {Tagliacozzo}, {Rosario}, {Veledina}, {Kim}, {Marin}, {Bianchi}, {Poutanen}, {Matt}, {Marshall}, {Ursini}, {De Rosa}, {Petrucci}, {Madejski}, {Barnouin}, {Gesu}, {Dov{\v{c}}iak}, {Gianolli}, {Krawczynski}, {Loktev}, {Middei}, {Podgorny}, {Puccetti}, {Ratheesh}, {Soffitta}, {Tombesi}, {Ehlert}, {Massaro}, {Agudo}, {Antonelli}, {Bachetti}, {Baldini}, {Baumgartner}, {Bellazzini}, {Bongiorno}, {Bonino}, {Brez}, {Bucciantini}, {Capitanio}, {Castellano}, {Cavazzuti}, {Chen}, {Ciprini}, {Costa}, {Del Monte}, {Lalla}, {Marco}, {Donnarumma}, {Doroshenko}, {Enoto}, {Evangelista}, {Fabiani}, {Ferrazzoli}, {Garc{\'\i}a}, {Gunji}, {Heyl}, {Iwakiri}, {Jorstad}, {Kaaret}, {Karas}, {Kislat}, {Kitaguchi}, {Kolodziejczak}, {Monaca}, {Latronico}, {Liodakis}, {Maldera}, {Manfreda}, {Marscher}, {Mitsuishi}, {Mizuno}, {Muleri}, {Negro}, {Ng}, {O'Dell}, {Omodei}, {Oppedisano}, {Papitto}, {Pavlov}, {Peirson}, {Perri}, {Pesce-Rollins}, {Pilia}, {Possenti}, {Ramsey},
  {Rankin}, {Roberts}, {Romani}, {Sgr{\`o}}, {Slane}, {Spandre}, {Swartz}, {Tamagawa}, {Tavecchio}, {Taverna}, {Tawara}, {Tennant}, {Thomas}, {Trois}, {Tsygankov}, {Turolla}, {Vink}, {Weisskopf}, {Wu}, {Xie}, \& {Zane}}]{Ingram2023}
{Ingram}, A., {Ewing}, M., {Marinucci}, A., {et~al.} 2023, \href{http://dx.doi.org/10.1093/mnras/stad2625}{\JournalTitle{\mnras}, 525, 5437}

\bibitem[{{Ingram} {et~al.}(2024){Ingram}, {Bollemeijer}, {Veledina}, {Dovciak}, {Poutanen}, {Egron}, {Russell}, {Trushkin}, {Negro}, {Ratheesh}, {Capitanio}, {Connors}, {Neilsen}, {Kraus}, {Noemi Iacolina}, {Pellizzoni}, {Pilia}, {Carotenuto}, {Matt}, {Mastroserio}, {Kaaret}, {Bianchi}, {Garcia}, {Bachetti}, {Wu}, {Costa}, {Ewing}, {Kravtsov}, {Krawczynski}, {Loktev}, {Marinucci}, {Marra}, {Mikusincova}, {Nathan}, {Parra}, {Petrucci}, {Righini}, {Soffitta}, {Steiner}, {Svoboda}, {Tombesi}, {Tugliani}, {Ursini}, {Yang}, {Zane}, {Zhang}, {Agudo}, {Antonelli}, {Baldini}, {Baumgartner}, {Bellazzini}, {Bongiorno}, {Bonino}, {Brez}, {Bucciantini}, {Castellano}, {Cavazzuti}, {Chen}, {Ciprini}, {De Rosa}, {Del Monte}, {Di Gesu}, {Di Lalla}, {Di Marco}, {Donnarumma}, {Doroshenko}, {Ehlert}, {Enoto}, {Evangelista}, {Fabiani}, {Ferrazzoli}, {Gunji}, {Hayashida}, {Heyl}, {Iwakiri}, {Jorstad}, {Karas}, {Kislat}, {Kitaguchi}, {Kolodziejczak}, {La Monaca}, {Latronico}, {Liodakis}, {Maldera}, {Manfreda}, {Marin},
  {Marscher}, {Marshall}, {Massaro}, {Mitsuishi}, {Mizuno}, {Muleri}, {Ng}, {O'Dell}, {Omodei}, {Oppedisano}, {Papitto}, {Pavlov}, {Peirson}, {Perri}, {Pesce-Rollins}, {Possenti}, {Puccetti}, {Ramsey}, {Rankin}, {Roberts}, {Romani}, {Sgro}, {Slane}, {Spandre}, {Swartz}, {Tamagawa}, {Tavecchio}, {Taverna}, {Tawara}, {Tennant}, {Thomas}, {Trois}, {Tsygankov}, {Turolla}, {Vink}, {Weisskopf}, \& {Xie}}]{Ingram2024}
{Ingram}, A., {Bollemeijer}, N., {Veledina}, A., {et~al.} 2024, \href{http://dx.doi.org/10.48550/arXiv.2311.05497}{\JournalTitle{\apj, in press}, arXiv:2311.05497}

\bibitem[{{Kennea} \& {Swift Team}(2023)}]{GCN.34540}
{Kennea}, J.~A., \& {Swift Team}. 2023, \JournalTitle{GRB Coordinates Network}, 34540, 1

\bibitem[{{Kravtsov} {et~al.}(2023){Kravtsov}, {Nitindala}, {Veledina}, {Berdyugin}, {Poutanen}, \& {Piirola}}]{Atel16245}
{Kravtsov}, V., {Nitindala}, A.~P., {Veledina}, A., {et~al.} 2023, \JournalTitle{The Astronomer's Telegram}, 16245, 1

\bibitem[{{Krawczynski} \& {Beheshtipour}(2022)}]{Krawczynski2022b}
{Krawczynski}, H., \& {Beheshtipour}, B. 2022, \href{http://dx.doi.org/10.3847/1538-4357/ac7725}{\JournalTitle{\apj}, 934, 4}

\bibitem[{{Krawczynski} {et~al.}(2022){Krawczynski}, {Muleri}, {Dov{\v{c}}iak}, {Veledina}, {Rodriguez Cavero}, {Svoboda}, {Ingram}, {Matt}, {Garcia}, {Loktev}, {Negro}, {Poutanen}, {Kitaguchi}, {Podgorn{\'y}}, {Rankin}, {Zhang}, {Berdyugin}, {Berdyugina}, {Bianchi}, {Blinov}, {Capitanio}, {Di Lalla}, {Draghis}, {Fabiani}, {Kagitani}, {Kravtsov}, {Kiehlmann}, {Latronico}, {Lutovinov}, {Mandarakas}, {Marin}, {Marinucci}, {Miller}, {Mizuno}, {Molkov}, {Omodei}, {Petrucci}, {Ratheesh}, {Sakanoi}, {Semena}, {Skalidis}, {Soffitta}, {Tennant}, {Thalhammer}, {Tombesi}, {Weisskopf}, {Wilms}, {Zhang}, {Agudo}, {Antonelli}, {Bachetti}, {Baldini}, {Baumgartner}, {Bellazzini}, {Bongiorno}, {Bonino}, {Brez}, {Bucciantini}, {Castellano}, {Cavazzuti}, {Ciprini}, {Costa}, {De Rosa}, {Del Monte}, {Di Gesu}, {Di Marco}, {Donnarumma}, {Doroshenko}, {Ehlert}, {Enoto}, {Evangelista}, {Ferrazzoli}, {Gunji}, {Hayashida}, {Heyl}, {Iwakiri}, {Jorstad}, {Karas}, {Kolodziejczak}, {La Monaca}, {Liodakis}, {Maldera}, {Manfreda},
  {Marscher}, {Marshall}, {Mitsuishi}, {Ng}, {O{\textquoteright}Dell}, {Oppedisano}, {Papitto}, {Pavlov}, {Peirson}, {Perri}, {Pesce-Rollins}, {Pilia}, {Possenti}, {Puccetti}, {Ramsey}, {Romani}, {Sgr{\`o}}, {Slane}, {Spandre}, {Tamagawa}, {Tavecchio}, {Taverna}, {Tawara}, {Thomas}, {Trois}, {Tsygankov}, {Turolla}, {Vink}, {Wu}, {Xie}, \& {Zane}}]{Krawczynski2022}
{Krawczynski}, H., {Muleri}, F., {Dov{\v{c}}iak}, M., {et~al.} 2022, \href{http://dx.doi.org/10.1126/science.add5399}{\JournalTitle{Science}, 378, 650}

\bibitem[{{Liu} {et~al.}(2023){Liu}, {Li}, {Pan}, {Liu}, {Ling}, {Zhang}, {Cheng}, {Cui}, {Fan}, {Hu}, {Hu}, {Huang}, {Jin}, {Li}, {Li}, {Li}, {Liu}, {Sun}, {Wang}, {Wang}, {Wang}, {Wu}, {Xu}, {Xu}, {Yang}, {Zhang}, {Zhang}, {Zhang}, {Zhang}, {Zhao}, \& {Yuan}}]{ATel16210}
{Liu}, H.~Y., {Li}, D.~Y., {Pan}, H.~W., {et~al.} 2023, \JournalTitle{The Astronomer's Telegram}, 16210, 1

\bibitem[{{Mata S{\'a}nchez} {et~al.}(2024){Mata S{\'a}nchez}, {Mu{\~n}oz-Darias}, {Armas Padilla}, {Casares}, \& {Torres}}]{Mata2024}
{Mata S{\'a}nchez}, D., {Mu{\~n}oz-Darias}, T., {Armas Padilla}, M., {Casares}, J., \& {Torres}, M.~A.~P. 2024, \href{http://dx.doi.org/10.1051/0004-6361/202348754}{\JournalTitle{\aap}, 682, L1}

\bibitem[{{Matsuoka} {et~al.}(2009){Matsuoka}, {Kawasaki}, {Ueno}, {Tomida}, {Kohama}, {Suzuki}, {Adachi}, {Ishikawa}, {Mihara}, {Sugizaki}, {Isobe}, {Nakagawa}, {Tsunemi}, {Miyata}, {Kawai}, {Kataoka}, {Morii}, {Yoshida}, {Negoro}, {Nakajima}, {Ueda}, {Chujo}, {Yamaoka}, {Yamazaki}, {Nakahira}, {You}, {Ishiwata}, {Miyoshi}, {Eguchi}, {Hiroi}, {Katayama}, \& {Ebisawa}}]{Maxi_2009}
{Matsuoka}, M., {Kawasaki}, K., {Ueno}, S., {et~al.} 2009, \href{http://dx.doi.org/10.1093/pasj/61.5.999}{\JournalTitle{\pasj}, 61, 999}

\bibitem[{{Matt}(1993)}]{Matt1993}
{Matt}, G. 1993, \href{http://dx.doi.org/10.1093/mnras/260.3.663}{\JournalTitle{\mnras}, 260, 663}

\bibitem[{{Miller-Jones} {et~al.}(2023){Miller-Jones}, {Sivakoff}, {Bahramian}, \& {Russell}}]{ATel16211}
{Miller-Jones}, J.~C.~A., {Sivakoff}, G.~R., {Bahramian}, A., \& {Russell}, T.~D. 2023, \JournalTitle{The Astronomer's Telegram}, 16211, 1

\bibitem[{{Narayan} \& {Yi}(1995)}]{NarayanYi95}
{Narayan}, R., \& {Yi}, I. 1995, \href{http://dx.doi.org/10.1086/176343}{\JournalTitle{\apj}, 452, 710}

\bibitem[{{Negoro} {et~al.}(2023){Negoro}, {Serino}, {Nakajima}, {Kobayashi}, {Tanaka}, {Soejima}, {Kudo}, {Mihara}, {Kawamuro}, {Yamada}, {Tamagawa}, {Kawai}, {Matsuoka}, {Sakamoto}, {Sugita}, {Hiramatsu}, {Nishikawa}, {Yoshida}, {Tsuboi}, {Urabe}, {Nawa}, {Nemoto}, {Shidatsu}, {Takahashi}, {Niwano}, {Sato}, {Higuchi}, {Yatsu}, {Nakahira}, {Ueno}, {Tomida}, {Ishikawa}, {Ogawa}, {Kurihara}, {Ueda}, {Setoguchi}, {Yoshitake}, {Nakatani}, {Yamauchi}, {Hagiwara}, {Umeki}, {Otsuki}, {Yamaoka}, {Kawakubo}, {Sugizaki}, {Iwakiri}, \& {MAXI Team}}]{GCN.34544}
{Negoro}, H., {Serino}, M., {Nakajima}, M., {et~al.} 2023, \JournalTitle{GRB Coordinates Network}, 34544, 1

\bibitem[{{Novikov} \& {Thorne}(1973)}]{Novikov1973}
{Novikov}, I.~D., \& {Thorne}, K.~S. 1973, in Black Holes (Les Astres Occlus) (New York: Gordon \& Breach), 343

\bibitem[{{Palmer} \& {Parsotan}(2023)}]{ATel16215}
{Palmer}, D.~M., \& {Parsotan}, T.~M. 2023, \JournalTitle{The Astronomer's Telegram}, 16215, 1

\bibitem[{{Peng} {et~al.}(2024){Peng}, {Zhang}, {Shui}, {Zhang}, {Kong}, {Chen}, {Wang}, {Ji}, {Qu}, {Tao}, {Ge}, {Chang}, {Li}, {Li}, {Yu}, \& {Yan}}]{Peng2024}
{Peng}, J.-Q., {Zhang}, S., {Shui}, Q.-C., {et~al.} 2024, \href{http://dx.doi.org/10.3847/2041-8213/ad17ca}{\JournalTitle{\apjl}, 960, L17}

\bibitem[{{Podgorný} {et~al.}(2024){Podgorný}, {Svoboda}, \& {Dovčiak}}]{ATel16541}
{Podgorný}, J., {Svoboda}, J., \& {Dovčiak}, M. 2024, \JournalTitle{The Astronomer's Telegram}, 16541, 1

\bibitem[{{Poutanen}(1994)}]{Poutanen1994}
{Poutanen}, J. 1994, \href{http://dx.doi.org/10.1086/192024}{\JournalTitle{\apjs}, 92, 607}

\bibitem[{{Poutanen} \& {Svensson}(1996)}]{Poutanen1996}
{Poutanen}, J., \& {Svensson}, R. 1996, \href{http://dx.doi.org/10.1086/177865}{\JournalTitle{\apj}, 470, 249}

\bibitem[{{Poutanen} \& {Veledina}(2014)}]{PoutanenVeledina2014}
{Poutanen}, J., \& {Veledina}, A. 2014, \href{http://dx.doi.org/10.1007/s11214-013-0033-3}{\JournalTitle{\ssr}, 183, 61}

\bibitem[{{Poutanen} {et~al.}(2023){Poutanen}, {Veledina}, \& {Beloborodov}}]{Poutanen2023}
{Poutanen}, J., {Veledina}, A., \& {Beloborodov}, A.~M. 2023, \href{http://dx.doi.org/10.3847/2041-8213/acd33e}{\JournalTitle{\apjl}, 949, L10}

\bibitem[{{Rees}(1975)}]{Rees1975}
{Rees}, M.~J. 1975, \href{http://dx.doi.org/10.1093/mnras/171.3.457}{\JournalTitle{\mnras}, 171, 457}

\bibitem[{{Remillard} \& {McClintock}(2006)}]{Remillard2006}
{Remillard}, R.~A., \& {McClintock}, J.~E. 2006, \href{http://dx.doi.org/10.1146/annurev.astro.44.051905.092532}{\JournalTitle{\araa}, 44, 49}

\bibitem[{{Russell} {et~al.}(2024){Russell}, {Carotenuto}, {Miller-Jones}, {Atri}, {Grollimund}, {Corbel}, \& {et al.}}]{ATel16552}
{Russell}, T.~D., {Carotenuto}, F., {Miller-Jones}, J.~C.~A., {et~al.} 2024, \JournalTitle{The Astronomer's Telegram}, 16552, 1

\bibitem[{{Shakura} \& {Sunyaev}(1973)}]{Shakura1973}
{Shakura}, N.~I., \& {Sunyaev}, R.~A. 1973, \JournalTitle{\aap}, 500, 33

\bibitem[{{Sunyaev} \& {Titarchuk}(1980)}]{Sunyaev1980}
{Sunyaev}, R.~A., \& {Titarchuk}, L.~G. 1980, \JournalTitle{\aap}, 86, 121

\bibitem[{{Sunyaev} \& {Titarchuk}(1985)}]{SunyaevTitarchuk1985}
---. 1985, \JournalTitle{\aap}, 143, 374

\bibitem[{{Sunyaev} {et~al.}(2023){Sunyaev}, {Mereminskiy}, {Molkov}, {Semena}, {Arefiev}, {Krivonos}, {Levin}, {Lutovinov}, {Shtykovsky}, \& {Tkachenko}}]{ATel16217}
{Sunyaev}, R.~A., {Mereminskiy}, I.~A., {Molkov}, S.~V., {et~al.} 2023, \JournalTitle{The Astronomer's Telegram}, 16217, 1

\bibitem[{{Svoboda} {et~al.}(2024){Svoboda}, {Dov{\v{c}}iak}, {Steiner}, {Kaaret}, {Podgorn{\'y}}, {Poutanen}, {Veledina}, {Muleri}, {Taverna}, {Krawczynski}, {Brigitte}, {Datta}, {Bianchi}, {Mu{\~n}oz-Darias}, {Negro}, {Rodriguez Cavero}, {Castro Segura}, {Bollemeijer}, {Garc{\'\i}a}, {Ingram}, {Matt}, {Nathan}, {Weisskopf}, {Altamirano}, {Baldini}, {Capitanio}, {Egron}, {Emami}, {Hu}, {Marra}, {Mastroserio}, {Petrucci}, {Ratheesh}, {Soffitta}, {Tombesi}, {Yang}, \& {Zhang}}]{Svoboda2024b}
{Svoboda}, J., {Dov{\v{c}}iak}, M., {Steiner}, J.~F., {et~al.} 2024, \href{http://dx.doi.org/10.3847/2041-8213/ad402e}{\JournalTitle{\apjl}, 966, L35}

\bibitem[{{Trushkin} {et~al.}(2023){Trushkin}, {Bursov}, {Nizhelskij}, \& {Tsybulev}}]{Atel16289}
{Trushkin}, S.~A., {Bursov}, N.~N., {Nizhelskij}, N.~A., \& {Tsybulev}, P.~G. 2023, \JournalTitle{The Astronomer's Telegram}, 16289, 1

\bibitem[{{Ursini} {et~al.}(2022){Ursini}, {Matt}, {Bianchi}, {Marinucci}, {Dov{\v{c}}iak}, \& {Zhang}}]{Ursini2022}
{Ursini}, F., {Matt}, G., {Bianchi}, S., {et~al.} 2022, \href{http://dx.doi.org/10.1093/mnras/stab3745}{\JournalTitle{\mnras}, 510, 3674}

\bibitem[{Veledina \& Poutanen(2022)}]{veledina22}
Veledina, A., \& Poutanen, J. 2022, {Polarization of Comptonized emission in slab geometry}

\bibitem[{{Veledina} {et~al.}(2013){Veledina}, {Poutanen}, \& {Vurm}}]{Veledina2013}
{Veledina}, A., {Poutanen}, J., \& {Vurm}, I. 2013, \href{http://dx.doi.org/10.1093/mnras/stt124}{\JournalTitle{\mnras}, 430, 3196}

\bibitem[{{Veledina} {et~al.}(2023){Veledina}, {Muleri}, {Dov{\v{c}}iak}, {Poutanen}, {Ratheesh}, {Capitanio}, {Matt}, {Soffitta}, {Tennant}, {Negro}, {Kaaret}, {Costa}, {Ingram}, {Svoboda}, {Krawczynski}, {Bianchi}, {Steiner}, {Garc{\'\i}a}, {Kravtsov}, {Nitindala}, {Ewing}, {Mastroserio}, {Marinucci}, {Ursini}, {Tombesi}, {Tsygankov}, {Yang}, {Weisskopf}, {Trushkin}, {Egron}, {Iacolina}, {Pilia}, {Marra}, {Miku{\v{s}}incov{\'a}}, {Nathan}, {Parra}, {Petrucci}, {Podgorn{\'y}}, {Tugliani}, {Zane}, {Zhang}, {Agudo}, {Antonelli}, {Bachetti}, {Baldini}, {Baumgartner}, {Bellazzini}, {Bongiorno}, {Bonino}, {Brez}, {Bucciantini}, {Castellano}, {Cavazzuti}, {Chen}, {Ciprini}, {De Rosa}, {Del Monte}, {Di Gesu}, {Di Lalla}, {Di Marco}, {Donnarumma}, {Doroshenko}, {Ehlert}, {Enoto}, {Evangelista}, {Fabiani}, {Ferrazzoli}, {Gunji}, {Hayashida}, {Heyl}, {Iwakiri}, {Jorstad}, {Karas}, {Kislat}, {Kitaguchi}, {Kolodziejczak}, {La Monaca}, {Latronico}, {Liodakis}, {Maldera}, {Manfreda}, {Marin}, {Marscher}, {Marshall},
  {Massaro}, {Mitsuishi}, {Mizuno}, {Ng}, {O'Dell}, {Omodei}, {Oppedisano}, {Papitto}, {Pavlov}, {Peirson}, {Perri}, {Pesce-Rollins}, {Possenti}, {Puccetti}, {Ramsey}, {Rankin}, {Roberts}, {Romani}, {Sgr{\`o}}, {Slane}, {Spandre}, {Swartz}, {Tamagawa}, {Tavecchio}, {Taverna}, {Tawara}, {Thomas}, {Trois}, {Turolla}, {Vink}, {Wu}, \& {Xie}}]{Veledina2023}
{Veledina}, A., {Muleri}, F., {Dov{\v{c}}iak}, M., {et~al.} 2023, \href{http://dx.doi.org/10.3847/2041-8213/ad0781}{\JournalTitle{\apjl}, 958, L16}

\bibitem[{{Vrtilek} {et~al.}(2023){Vrtilek}, {Gurwell}, {McCollough}, \& {Rao}}]{ATel16230}
{Vrtilek}, S.~D., {Gurwell}, M., {McCollough}, M., \& {Rao}, R. 2023, \JournalTitle{The Astronomer's Telegram}, 16230, 1

\bibitem[{{Weisskopf} {et~al.}(2022){Weisskopf}, {Soffitta}, {Baldini}, {Ramsey}, {O'Dell}, {Romani}, {Matt}, {Deininger}, {Baumgartner}, {Bellazzini}, {Costa}, {Kolodziejczak}, {Latronico}, {Marshall}, {Muleri}, {Bongiorno}, {Tennant}, {Bucciantini}, {Dovciak}, {Marin}, {Marscher}, {Poutanen}, {Slane}, {Turolla}, {Kalinowski}, {Di Marco}, {Fabiani}, {Minuti}, {La Monaca}, {Pinchera}, {Rankin}, {Sgro'}, {Trois}, {Xie}, {Alexander}, {Allen}, {Amici}, {Andersen}, {Antonelli}, {Antoniak}, {Attina'}, {Barbanera}, {Bachetti}, {Baggett}, {Bladt}, {Brez}, {Bonino}, {Boree}, {Borotto}, {Breeding}, {Brienza}, {Bygott}, {Caporale}, {Cardelli}, {Carpentiero}, {Castellano}, {Castronuovo}, {Cavalli}, {Cavazzuti}, {Ceccanti}, {Centrone}, {Citraro}, {D'Amico}, {D'Alba}, {Di Gesu}, {Del Monte}, {Dietz}, {Di Lalla}, {Di Persio}, {Dolan}, {Donnarumma}, {Evangelista}, {Ferrant}, {Ferrazzoli}, {Ferrie}, {Footdale}, {Forsyth}, {Foster}, {Garelick}, {Gunji}, {Gurnee}, {Head}, {Hibbard}, {Johnson}, {Kelly}, {Kilaru}, {Lefevre}, {Le
  Roy}, {Loffredo}, {Lorenzi}, {Lucchesi}, {Maddox}, {Magazzu}, {Maldera}, {Manfreda}, {Mangraviti}, {Marengo}, {Marrocchesi}, {Massaro}, {Mauger}, {McCracken}, {McEachen}, {Mize}, {Mereu}, {Mitchell}, {Mitsuishi}, {Morbidini}, {Mosti}, {Nasimi}, {Negri}, {Negro}, {Nguyen}, {Nitschke}, {Nuti}, {Onizuka}, {Oppedisano}, {Orsini}, {Osborne}, {Pacheco}, {Paggi}, {Painter}, {Pavelitz}, {Pentz}, {Piazzolla}, {Perri}, {Pesce-Rollins}, {Peterson}, {Pilia}, {Profeti}, {Puccetti}, {Ranganathan}, {Ratheesh}, {Reedy}, {Root}, {Rubini}, {Ruswick}, {Sanchez}, {Sarra}, {Santoli}, {Scalise}, {Sciortino}, {Schroeder}, {Seek}, {Sosdian}, {Spandre}, {Speegle}, {Tamagawa}, {Tardiola}, {Tobia}, {Thomas}, {Valerie}, {Vimercati}, {Walden}, {Weddendorf}, {Wedmore}, {Welch}, {Zanetti}, \& {Zanetti}}]{Weisskopf2022}
{Weisskopf}, M.~C., {Soffitta}, P., {Baldini}, L., {et~al.} 2022, \href{http://dx.doi.org/10.1117/1.JATIS.8.2.026002}{\JournalTitle{JATIS}, 8, 026002}

\bibitem[{{Wilms} {et~al.}(2000){Wilms}, {Allen}, \& {McCray}}]{Wilms2000}
{Wilms}, J., {Allen}, A., \& {McCray}, R. 2000, \href{http://dx.doi.org/10.1086/317016}{\JournalTitle{\apj}, 542, 914}

\bibitem[{{Yuan} \& {Narayan}(2014)}]{YuanNarayan2014}
{Yuan}, F., \& {Narayan}, R. 2014, \href{http://dx.doi.org/10.1146/annurev-astro-082812-141003}{\JournalTitle{\araa}, 52, 529}

\bibitem[{{Zdziarski} \& {Gierli{\'n}ski}(2004)}]{Zdziarski2004}
{Zdziarski}, A.~A., \& {Gierli{\'n}ski}, M. 2004, \href{http://dx.doi.org/10.1143/PTPS.155.99}{\JournalTitle{Progr. Theor. Phys. Suppl.}, 155, 99}

\bibitem[{{Zhao} {et~al.}(2024){Zhao}, {Tao}, {Li}, {Zhang}, {Feng}, {Ge}, {Ji}, {Wang}, {Huang}, {Ma}, {Zhang}, {Qu}, {Xu}, {Zhang}, {Yin}, {Shui}, {Ma}, {Zhao}, {Li}, {Yang}, {Liu}, \& {Yu}}]{Zhao2024}
{Zhao}, Q.-C., {Tao}, L., {Li}, H.-C., {et~al.} 2024, \href{http://dx.doi.org/10.3847/2041-8213/ad1e6c}{\JournalTitle{\apjl}, 961, L42}

\end{thebibliography}
\bibliographystyle{yahapj}

\begin{appendix}
\section{Time analysis}\label{timing_analysis}

\begin{figure} 
\centering
\includegraphics[width=\linewidth]{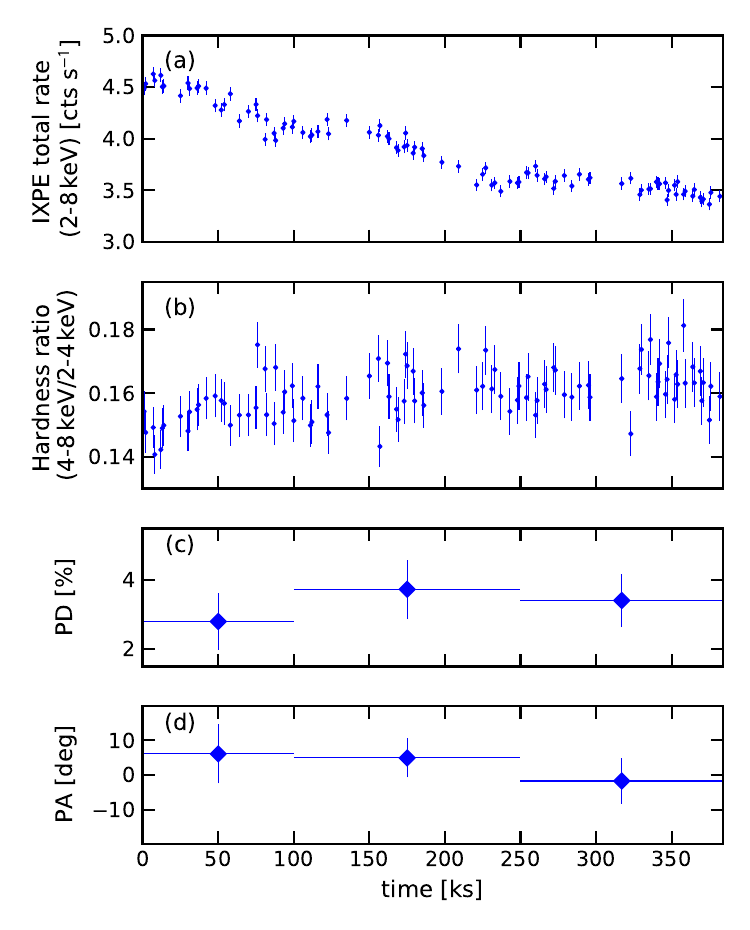}
\caption{Time analysis of the IXPE data. (a) IXPE light curve in the 2--8 keV band during observation 8 of Swift~J1727.8$-$1613 with a time bin of 1~ks. (b) Evolution of the IXPE hardness ratio between the 4--8 and 2--4 keV bands. (c) and (d):  2--8 keV PD and PA from IXPE data, respectively, determined using Model (\ref{model_polconst}) in {\sc xspec} in the corresponding restricted time ranges.} \label{fig:lightcurves}
\end{figure}

Figure \ref{fig:lightcurves} shows the IXPE light curve in 2--8 keV during observation 8 of Swift~J1727.8$-$1613. The count rate systematically decreased during this observation by about 30\%. We show in the same figure that the hardness ratio slightly increased in the meantime, as the source underwent the last phase of the soft-to-hard state transition.

To test whether the X-ray polarisation properties of the source changed between the beginning and the end of observation 8, we performed the same analysis in the 2--8 keV energy range as described in Sect.~\ref{sec:results} but for a restricted time range. Using Model (\ref{model_polconst}) in {\sc xspec}, we obtained $\chi^2 / {\rm dof} = 1383/1333$,  1396/1333, and 1428/1333 for the time bins 0--100, 100--250, and 250--384~ks since the beginning of observation 8, respectively. The best-fit spectral parameter values are reported in Table~\ref{t:timerestricted}. The resulting best-fit PD and PA  values are shown in Fig.~\ref{fig:lightcurves} for the corresponding time bins. The polarisation properties do not show any significant variation with time despite the nearly monotonically decreasing flux during the observation. However, to claim any spectro-polarimetric properties changing with time within a certain statistical confidence, a more detailed timing analysis is required, preferably using multiple observatories. Such analysis is beyond the scope of this Letter and will be done in a follow-up work.

\begin{table}
\begin{center}
\caption{Best-fit spectral parameter values of the joint IXPE spectro-polarimetric fit in {\sc xspec} with Model (\ref{model_polconst}) for the data restricted to 0--100, 100--250, and 250--384~ks after the start of observation~8.}
\resizebox{\columnwidth}{!}{
\begin{tabular}{p{1.5cm}p{2.3cm}p{1.9cm}p{1.9cm}p{1.9cm}}
\hline\hline 
Component & Parameter [unit] &  \multicolumn{3}{c}{Interval}\\ 
\cline{3-5}
  &   &   0--100 ks &  100--250 ks &  250--384 ks \\ 
\hline \hline
{\tt tbabs} & $N_\textrm{H}$ [$10^{22}$\,cm$^{-2}$] & $0.24$ (frozen) & $0.24$ (frozen) & $0.24$ (frozen) \\
\hline
{\tt diskbb} & $kT_\mathrm{in}$ [keV] & $0.38\pm 0.07$ & $0.31\pm 0.06$ & $0.24\substack{+0.12\\-0.09}$ \\
    & norm [$10^2$] & $2\substack{+5\\-1}$ & $10\substack{+39\\-7}$ & $28\substack{+455\\-25}$ \\
\hline
{\tt powerlaw} & $\Gamma$ &
$1.80\substack{+0.03\\-0.04}$ &
$1.75\substack{+0.02\\-0.03}$ &
$1.77\pm 0.02$ \\
& norm & $0.075\substack{+0.004\\-0.005}$ & $0.093\substack{+0.003\\-0.004}$ & $0.084\substack{+0.002\\-0.003}$ \\
\hline
{\tt const} & factor DU 1 & $1.0$ (frozen) & $1.0$ (frozen) & $1.0$ (frozen) \\
& factor DU 2 & $1.028\pm 0.005$ & $1.031\pm 0.004$ & $1.048\pm 0.005$\\
&  factor DU 3 & $1.014\pm 0.005$ & $1.020\pm 0.004$ & $1.032\pm 0.005$ \\
\hline
\end{tabular}
}
\label{t:timerestricted}
\end{center}
\end{table}

\end{appendix}

\end{document}